\documentclass{article}
\usepackage[utf8]{inputenc}

\usepackage{amsmath,amssymb}
\newtheorem{theorem}{Theorem}

\usepackage{xcolor}
\usepackage{algorithm}
\usepackage{algpseudocode}
\usepackage{array}

\usepackage[english]{babel}

\usepackage{graphicx}

\title{Spatial Dynamics Behavioral Analysis of Motivational Operations Using Weighted Voronoi Diagrams}

\author{
Carlos Alberto Hernández-Linares\thanks{Universidad Veracruzana, Facultad de Matemáticas, Xalapa, Ver., México.} \and
Porfirio Toledo\footnotemark[1] \and
Brenda Zarahí Medina-Pérez\footnotemark[1] \and
Varsovia Hernández\thanks{Universidad Veracruzana, Centro de Investigaciones Biomédicas, Laboratorio de Psicología Comparada,  Xalapa, Ver., México.} \and 
Martha Lorena Avendaño-Garrido\footnotemark[1] \and
Víctor Quintero\footnotemark[2] \and
Alejandro León\footnotemark[2]
}
\date{}

\begin{document}

\maketitle

\begin{abstract}
This paper presents a novel approach to the analysis of spatial behavior distribution, utilizing weighted Voronoi diagrams. The objective is to map and understand how an experimental subject moves and spends time in various areas of a given space, thus identifying the areas of greatest behavioral interest. The technique entails the partitioning of the space into a grid, the designation of generator points, and the assignment of weights based on the time the subject spends in each region. The data analyzed were derived from multiple experimental sessions in which subjects were exposed to various conditions, including food deprivation, water deprivation, and combined deprivation and no deprivation. The aforementioned conditions resulted in the formation of clearly delineated spatial patterns. Weighted Voronoi diagrams provided a comprehensive and precise representation of these areas of interest, facilitating an in-depth examination of the evolution of behavioral patterns in diverse contexts, such as under different Motivational Operations. This tool offers a valuable perspective for the dynamic study of spatial behaviors in variable experimental settings.
\end{abstract}

\noindent
\textbf{Keywords.} 
Weighted Voronoi diagrams,
behavior analysis,
spatial dynamics behavioral analysis,
motivational operations

\section{Introduction}
\label{sec:introduction}

Spatial Dynamics Behavioral Analysis (SDBA) examines the movement patterns of individuals, whether animals or humans, concerning the functional organization of behavioral phenomena, including motivation, learning, and fear, among others \cite{leon2021}. This analysis aims to comprehend how spatiotemporal features of the environment and other variables, such as motivational states, interact and influence the spatial aspects of behavior. A significant challenge in SDBA is the effective visual representation of the quantitative relationships between such features and variables in a manner that is both clear and understandable. To this end, SDBA requires the implementation of methodologies that enable the partitioning of space into zones demarcated by the organism's own behavior, thus facilitating the analysis of Regions of Behavioral Relevance (RBR). These RBRs represent areas where significant behaviors occur, and their identification and tracking under different conditions are crucial for understanding the dynamics of these behaviors. In this study, we propose using weighted Voronoi diagrams as a potential solution for identifying RBRs, aiming to improve their representation and analysis under different conditions of deprivation: food, water, food-water, and no deprivation.

A Voronoi diagram is a mathematical construction that divides a given space into regions based on a finite set of points, called generators. Each region is associated with a single generator point, and each point in the space belongs to the region closest to one of the generators \cite{gallier2017,boots2009,preparata2012}. The incorporation of a weight associated with each generator point generates a weighted Voronoi diagram. This study focuses on the application of multiplicative weighting to Voronoi diagrams, where the regions are adjusted according to the weights assigned to the generators. This approach allows for a more accurate representation of the spatial and behavioral influences at any given point under each condition.

The application of Voronoi diagrams has been widespread in modeling and analyzing spatial problems. In geographic research \cite{mu2004}, resource allocation in urban areas \cite{Cui2021}, and urban revitalization and spatial visualization \cite{Figurska2022}, they have provided valuable tools for effective urban planning and the improvement of urban structure. In the context of fire monitoring via the use of drones, they have been demonstrated to be a valuable tool for predicting the spread of fire and optimizing patrol routes \cite{Giuseppi2021}. Furthermore, in the context of image reconstruction, the use of Voronoi diagrams based on natural neighbor interpolation has led to considerable improvements in image quality \cite{Enriquez-Cervantes2015}. Moreover, Voronoi diagrams have played a significant role in areas close to SDBA, particularly in the context of urban mobility. They have been instrumental in determining population concentration, identifying commuting patterns, and analyzing the speed and direction of movement \cite{Manca2017,Qu2021,Viloria2020}. In the field of sports science, Voronoi diagrams have been utilized to characterize collective behavior, identify significant moments in games, and analyze the spatial interaction between players \cite{Eliakim2022,Fonseca2013,Gudmundsson2018}.

The potential of weighted Voronoi diagrams in these contexts has not yet been fully exploited, thereby limiting their use to the classification of movements or the determination of relevant regions in individual episodes or single sessions more than contrasting the spatial organization of behavior as a function of given variables. This paper proposes the use of weighted Voronoi diagrams to analyze the behavioral dynamics associated with different Motivational Operations \cite{Lewon_2019,michael_establishing_1993} under various deprivation conditions. This approach allows for a functional analysis of the zones of influence under different deprivation conditions, emerging from the trajectories generated in each session. It is intended that this will provide a robust tool for the dynamic analysis of changes in the RBRs, offering a more comprehensive view of the spatial organization of behavior under different Motivational Operations.

\section{Weighted Voronoi diagrams} 

Consider a set of  generator points $P = \{p_1, p_2, \ldots, p_n\} \in \mathbb{R}^2$, which we will call the generator set, and a set of weights $W = \{w_1, w_2, \ldots, w_n\} \in \mathbb{R}^+$, where each parameter $w_i$ is the weight associated with the point $p_i$, which represents the ability of $p_i$ to influence the space. In order to reflect this influence, the value of $w_i$ is employed to define a weighted distance relative to $p_i$, which is denoted as $d_w(x, p_i)$. The specific manner in which $d_w$ is defined is contingent upon the type of weighted Voronoi diagram that is being considered.

For a given point $p_i \in P$, the weighted Voronoi polygon associated with this point is defined as follows:
\[ Vor(p_i) = \{ x\in \mathbb{R}^2 : d_w(x, p_i) \leq d_w(x, p_j), j\neq i \}. \]
The weighted Voronoi diagram generated by the set $P$ is a partition of $\mathbb{R}^2$ into $n$ regions defined by the Voronoi polygons associated with all $p_i \in P$, and is denoted by
\[ Vor(P) = \{ Vor(p_1), \ldots, Vor(p_n) \}.\]

In the case of two distinct points, $p_i, p_j \in P$, if the intersection of the polygons associated with these points contains more than a single point, it will be the weighted edge associated with $p_i$ and $p_j$, that is,
\[ e(p_i, p_j) = Vor(p_i) \cap Vor(p_j). \]
We will call the union of weighted Voronoi edges a Voronoi lattice. 
We also define the weighted bisector between these points as
\[ b(p_i, p_j) = \{ x \in \mathbb{R}^2 : d_w(x, p_i) = d_w(x, p_j) \}.\]

If we define the domain region of $p_i$ over $p_j$ as the set
\[ Dom(p_i, p_j) = \{ x \in \mathbb{R}^2 : d_w(x, p_i) \leq d_w(x, p_j), \}, \]
then we have
\[\begin{split} 
Vor(p_i) &= \bigcap_{j \neq i, j = 1}^n Dom(p_i, p_j), \\
b(p_i, p_j) &= Dom(p_i, p_j) \cap Dom(p_j, p_i). 
\end{split}\]
Thus, the bisector divides the space into the two domain regions $p_i$ and $p_j$ over $p_j$ and $p_i$, respectively.

There are several variants of weighted Voronoi diagrams, such as multiplicatively weighted, additively weighted, compoundly weighted, and power weighted. The choice of one variant or another depends on how the values of the set of weights $W$ are used to define the function $d_w$ (see \cite{boots2009}). In this study we will focus on multiplicatively weighted Voronoi diagrams, and thus the weighted distance will be considered as follows: 
\[d_{w}(x,p_i)= \frac{1}{w_i}\|x-p_i\|, \]
where $\|x - p_i\|$ is the Euclidean distance between points $x$ and $p_i$.

The shape of the resulting domains is contingent upon the relationship between the weights associated with two points, $p_i, p_j\in P$. In consideration of the weighted distance previously defined, if $w_i < w_j$, the domain region of point $p_i$ over point $p_j$ can be expressed as the closed ball:
\[Dom(p_i,p_j) = \left\lbrace x \in \mathbb{R}^2 : \| x - o \| \leq r \right\rbrace ,\]
where 
\[\begin{split}
o &= \frac{w_j^2}{w_j^2- w_i^2} p_i - \frac{w_i^2}{w_j^2- w_i^2} p_j, \\ 
r &= \frac{w_iw_j}{w_j^2-w_i^2}\|p_i-p_j\|.
\end{split}\]
Conversely, if $w_i > w_j$, the domain region of $p_i$ over $p_j$ is given by the following equation:
\[ Dom(p_i,p_j) = \left\lbrace x \in \mathbb{R}^2 : \| x - o \| \geq r \right\rbrace,\]
This equation defines the complement of the open ball with center at $o$ and radius $r$. When $w_i = w_j$, the domain region corresponds to the classical unweighted definition. Thus, we have the following result \cite{boots2009}. 

\begin{theorem}
The edges of a multiplicatively weighted Voronoi region are circular arcs if and only if the weights of two adjacent regions are not equal, and are straight lines in the plane if and only if the weights of two adjacent regions are equal. 
\end{theorem}

A variety of algorithms have been developed for the construction of Voronoi diagrams, including Domain Intersection, Incremental Algorithm, and Divide and Conquer method, among others  \cite{aurenhammer2013, deBerg2008, boots2009}. In this type of diagrams, the largest weights associated with the corresponding generating point determine more important regions and geometrically correspond to larger regions, with greater area.

\begin{figure}[h!]
\centering
\includegraphics[scale=0.35]{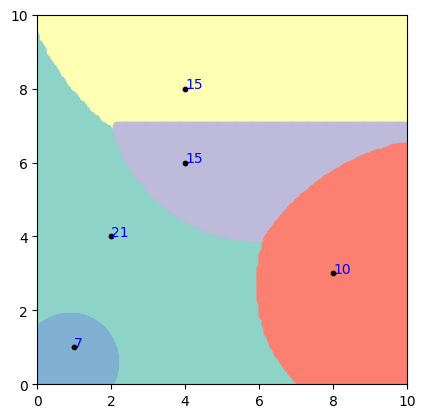}
\includegraphics[scale=0.35]{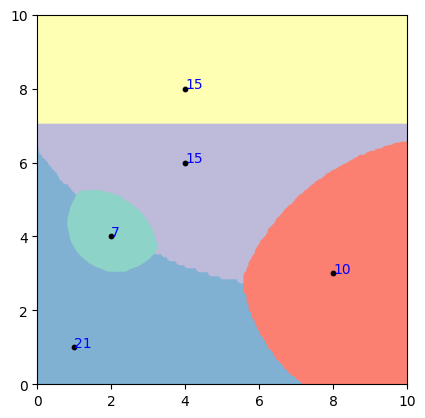}
\caption{Multiplicatively weighted Voronoi diagrams, the numbers represent the corresponding weights.}
\label{VoronoiDP}
\end{figure}

As illustrated in Fig. \ref{VoronoiDP}, two distinct Voronoi diagrams are presented for the same generator set, but with different sets of weights. It should be noted that the Voronoi regions are not necessarily convex.

\section{Application for SDBA}

Behavior Analysis (BA) examines the behavior of individuals by exploring the variables that influence it. BA studies the factors that affect behavior, traditionally focusing on single-discrete responses. However, studying organism movement has become crucial in understanding various behavioral phenomena within different ecological contexts. This approach is known as Spatial Dynamics Behavior Analysis (SDBA) (see \cite{leon2020ecological}).

SDBA can be applied to the study of specific behavioral phenomena, such as Motivational Operations (MOs), by analyzing the effect of different deprivation conditions (food, water, and combined food-and-water deprivation) on the emergence of different spatial dynamics when rats are given a choice between consuming water or food in a concurrent delivery arrangement. While SDBA represents an important advancement in BA, it presents several significant challenges, one of which is visualizing complex spatial behavioral patterns. For example, understanding how water or food deprivation affects the distribution of stays in specific zones where these resources are found is a primary focus.

To address this challenge, we will explore using weighted Voronoi diagrams in the context of MOs. First, we describe the experimental setting and method, followed by the data collected. Finally, we analyze the data, focusing on the spatial organization of behavior under different deprivation conditions using weighted Voronoi diagrams.

\subsection{Data Acquisition}

The study was conducted by members of the Comparative Psychology Laboratory at Universidad Veracruzana. The experimental subjects were six Wistar rats subjected to either a water or food restriction schedule, depending on the corresponding experimental phase, as described below.

A single-case experimental design was implemented following established behavior analysis protocols, incorporating a detailed behavioral record for each rat ($7200$ data points of the rat’s location per session). For data collection, we used a modified open field chamber with an experimental space measuring $92cm \times 92cm \times 33cm$ in length (width, and height). A limited-availability water dispenser was placed in the middle of one wall of the chamber at coordinates $(46, 0)$, and a limited-availability food dispenser was placed on the opposite wall at coordinates $(46, 92)$. We used a tracking system (EthoVision XT) to monitor the rats’ displacement inside the chamber throughout $20$ minutes in a potential food-or water- seeking behavior. The data provided indicate the subjects´ trajectory in coordinates $(x, y)$ over time, recorded at a resolution of $5$ frames per second.

Twenty-four sessions, each lasting $20$ minutes, were conducted under one of four deprivation conditions ($6$ sessions per condition), while the food and water delivery schedule during experimental sessions remained constant. Different deprivation conditions could involve different Motivational Operations (MOs). The deprivation conditions were: a) Water Deprivation (WD) b) Food Deprivation (FD) c) Food and Water Deprivation (FWD) d) No Deprivation (ND).

For each deprivation condition, food and/or water consumption was restricted for $22$ hours before each experimental session. Each deprivation condition consisted of three days with the corresponding restriction. Each experimental session consisted of concurrently presenting food and water in each dispenser every $30$ seconds, with limited availability of $3$ seconds. A yellow light above both dispensers was turned on with every delivery and remained on during the $3$ seconds of availability.

\subsection{Generation of weighted Voronoi diagrams}

In order to construct weighted Voronoi diagrams based on the collected experimental data, it is first necessary to determine a generator set and then assign appropriate weights to each of its points.

To this end, the experimental space was partitioned into a uniform grid of $n \times m$ regions, which defined an initial set $Q$ of $n \times m$ points, comprising the centers of these regions. The data obtained from the trajectory of the subject in each session within the experimental space were calculated using MOTUS software, allowing the determination of the time accumulated by the subject in each region during the session \cite{leon2020motus}. This information was then integrated into a matrix of size $n \times m$. The generator set, denoted by $P$, was defined as the set of points in $Q$ corresponding to the regions with nonzero cumulative time. The set of weights, represented by $W$, consists of positive cumulative times in these regions. Finally, the set $P$ is ordered in descending order with respect to the weight associated with its points.

The Algorithm~\ref{alg:MWP} illustrates the method for obtaining the Voronoi region associated with a given point. The functions $Dom(p_i, p_k)$ and $InArea(point, D)$ perform the following operations: the function $Dom(p_i, p_k)$ identifies the domain region of the generator point $p_i$ over $p_k$, while the function $InArea(point, D)$ determines whether a point belongs to $Vor(p_i)$. 

\begin{algorithm}[h!]
\caption{Voronoi Diagram}\label{alg:MWP}
    \begin{algorithmic}[0]
        \Require{$P$ (generator set), $W$ (weights set) and $cl$ (color set)}
        \Ensure{Voronoi regions}
        \For{$p_i \in P$ and $w_i \in W$}
            \If{$p_i = p_0$}
                \State{paint the background with color $cl[p_i]$}
            \Else
                \State{$D \gets []$ \Comment{empty list}}
                \For{$p_k\in P$}
                    \If{$p_k \neq p_i$}
                    \State{add $Dom(p_i, p_k)$ to $D$} \Comment{calculate the domain of the region}
                    \EndIf
                \EndFor
            \EndIf
            \State{$c \gets []$ \Comment{empty list}}
            \For{point in $Dom(p_i, p_0)$}
                \State{$InArea(point,D)$ add to $c$}
            \EndFor
            \State{paint $c$ with color $cl[p_i]$}
        \EndFor
        \State{draw the points of $P$}
    \end{algorithmic}
\end{algorithm}

The Algorithm~\ref{alg:MWP} can be summarized as follows: Given that the generating points are ordered in descending order of weight, the $Vor(p_1)$ region will be the largest, thus determining the background color. Subsequently, for a specific generator point $p_i \neq p_1$, all domains $Dom(p_i,p_j)$ for $i\neq j$ are saved in a list $D$. Thereafter, the set of points that are in all domains of the list $D$, that is, all points in $Vor(p_i)$, is constructed using the $InArea(point,D)$ function. The aforementioned points are then stored in a list, designated as $c$, along with the corresponding color assigned to the $Vor(p_i)$ region. The points are subsequently rendered. At the conclusion of this process, the generating points are then rendered. This approach allows us to paint over the previous color layer. In the absence of an ordering of the points, the generator with the highest weight would eventually cover the Voronoi regions of the preceding generators.
\begin{figure}[h!]
\centering
\includegraphics[width=0.235\textwidth]{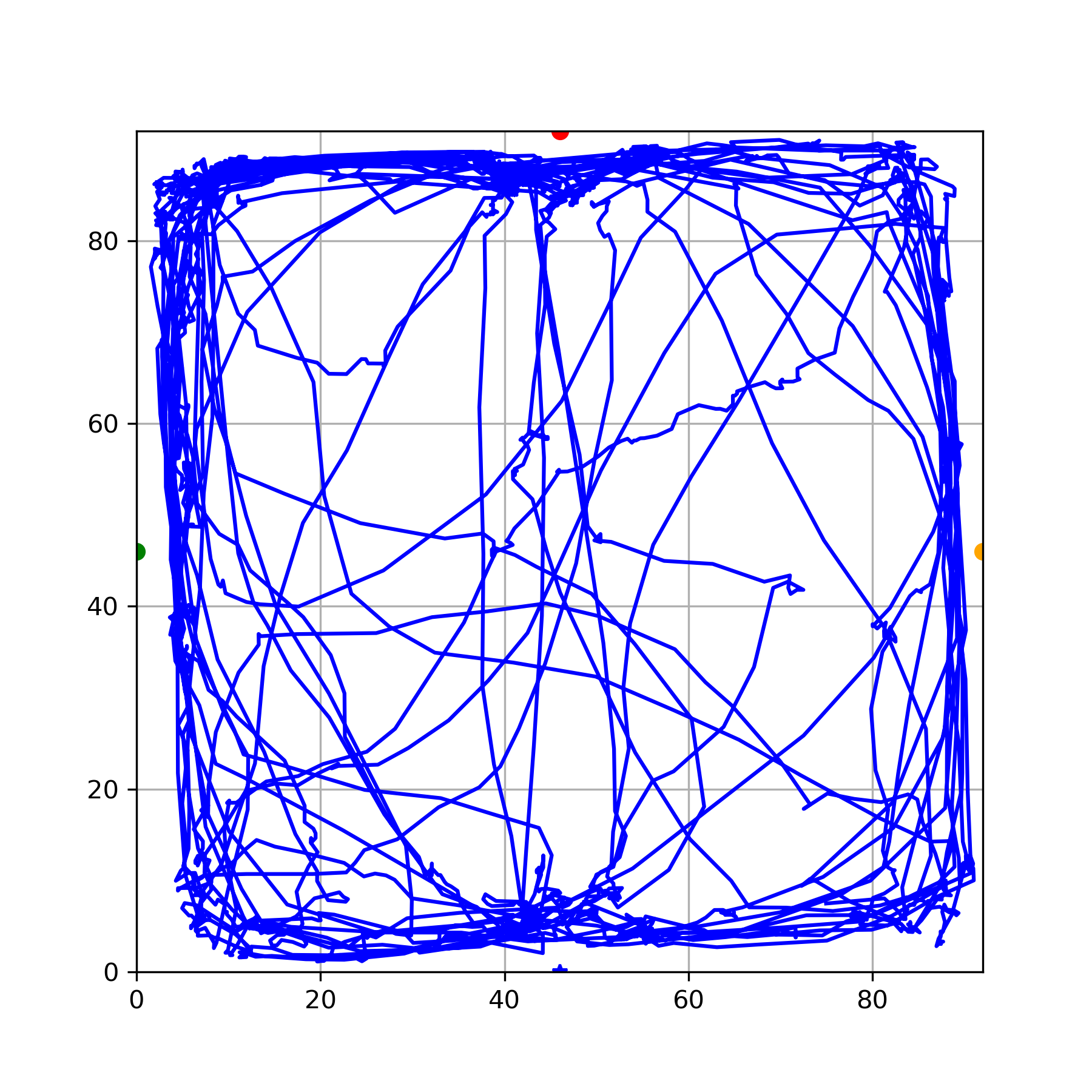}  
\includegraphics[width=0.228\textwidth]{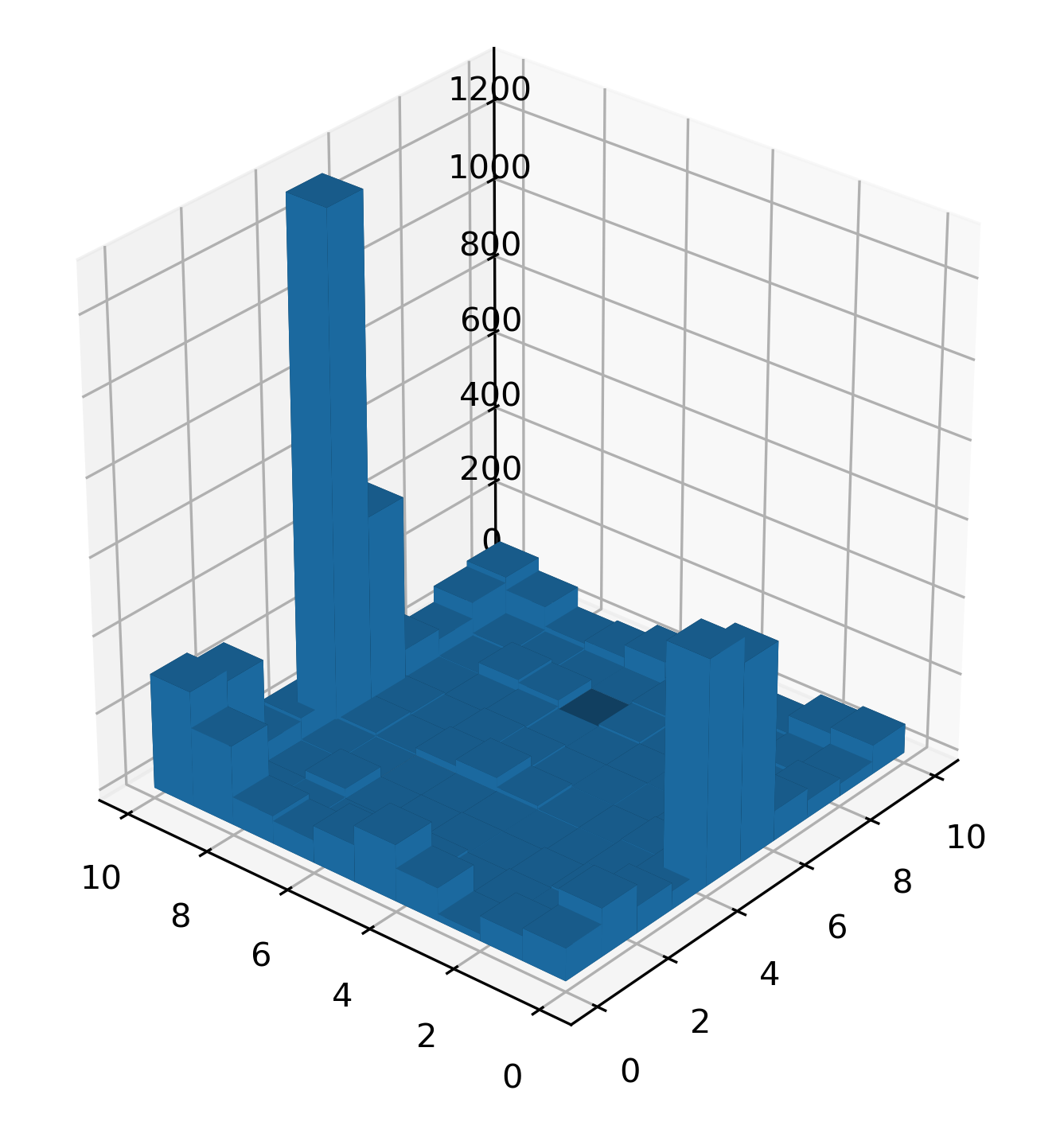}
\includegraphics[width=0.223\textwidth]{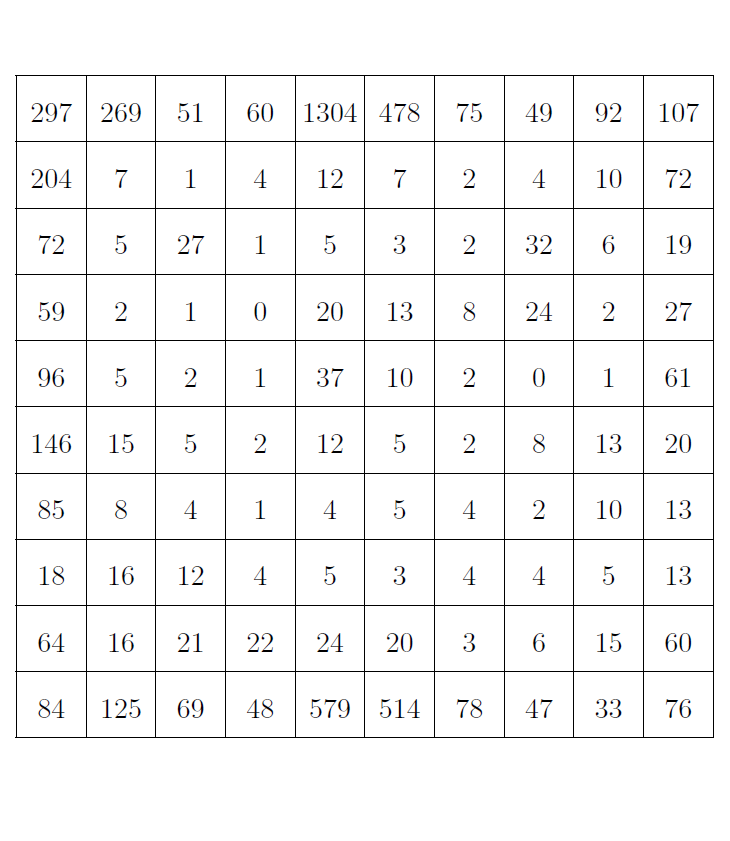}
\includegraphics[width=0.231\textwidth]{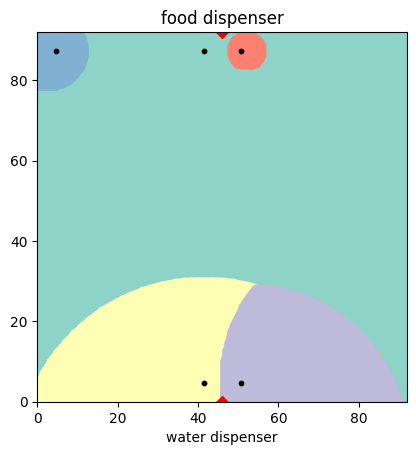}
\caption{Trajectory, accumulated time in regions and the corresponding multiplicatively weighted Voronoi diagram.}
\label{fig:Regi}
\end{figure}

Fig. \ref{fig:Regi} provides a visual representation of the process described above. In this case, the $92 cm \times 92 cm$ experimental space was divided into a $10 \times 10$ grid, generating an initial set of $100$ points. From the trajectory of the subject during the course of the session, the generator set and the associated set of weights were determined. As illustrated in Fig. \ref{fig:Regi}, the multiplicatively weighted Voronoi diagram features red triangles indicating the locations of the water dispensers and colored regions representing the weighted Voronoi regions of the generator points, marked with black dots. It should be noted that, for the purposes of simplifying the visualisation of the regions identified in this study, only the generator points with the five highest weights were included.

\subsection{Data Analysis}

\begin{figure}[h!]
\centering
\begin{tabular}{w{c}{0.6cm}m{3.07cm}m{3.07cm}}
& \hspace{0.8cm}{\footnotesize Subject 1} & \hspace{0.7cm}{\footnotesize Subject 5} \\
{\scriptsize FD} & 
\hspace{-0.3cm}\includegraphics[width=0.2\textwidth]{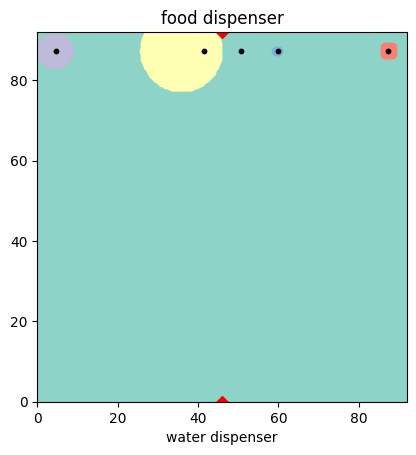} &
\hspace{-0.4cm}\includegraphics[width=0.2\textwidth]{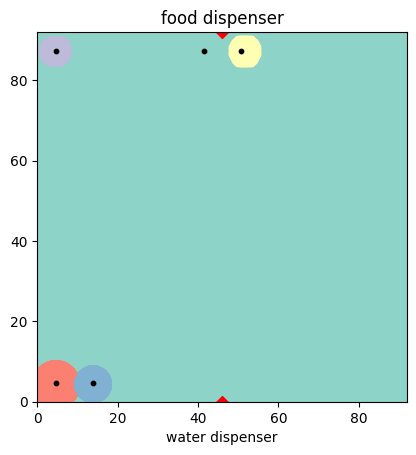} \\
{\scriptsize WD} &
\hspace{-0.3cm}\includegraphics[width=0.2\textwidth]{Figures/Suj1_s1_PA.png} &
\hspace{-0.4cm}\includegraphics[width=0.2\textwidth]{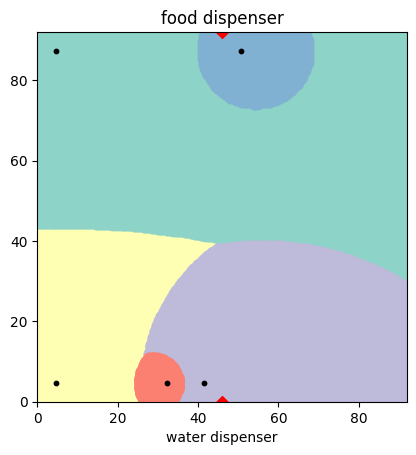} \\
{\scriptsize FWD} & 
\hspace{-0.3cm}\includegraphics[width=0.2\textwidth]{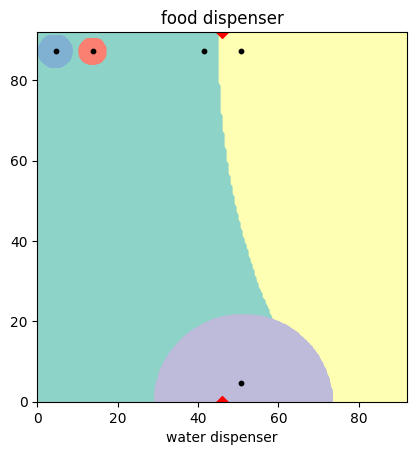} &
\hspace{-0.4cm}\includegraphics[width=0.2\textwidth]{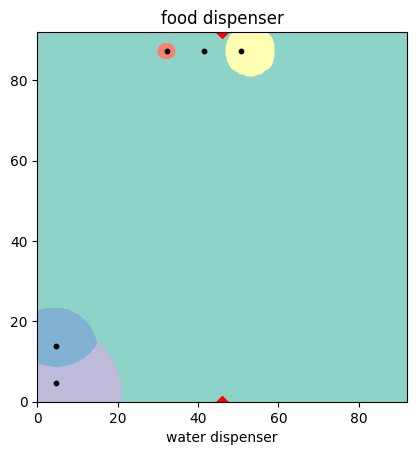} \\
{\scriptsize ND} &
\hspace{-0.3cm}\includegraphics[width=0.2\textwidth]{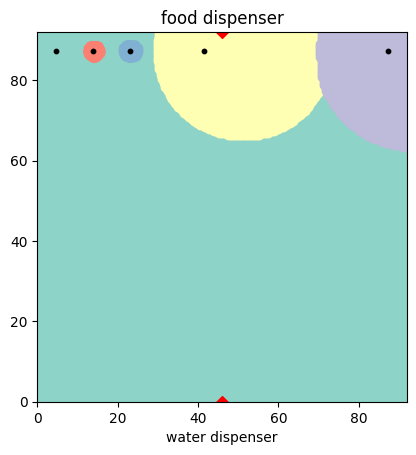} &
\hspace{-0.4cm}\includegraphics[width=0.2\textwidth]{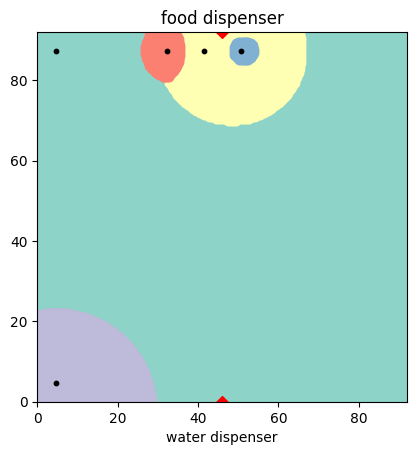} \\
\end{tabular}
\caption{Voronoi diagrams for subjects 1 and 5, illustrating a given session under conditions of Food Deprivation (FD), Water Deprivation (WD), Food and Water Deprivation (FWD), and No Deprivation (ND).}
\label{fig:subject1}
\end{figure}

To illustrate the use of Voronoi diagrams in SDBA, two subjects were selected from the six experimental subjects, with each subject exposed to the four aforementioned conditions. Fig. \ref{fig:subject1} display the subject and four sessions, one for each deprivation condition, in which the weighted Voronoi diagram of the five generators with the highest weight was generated. The Voronoi diagrams show the behavioral segmentation of the experimental space. The sessions were chosen to depict the differences on behavioral spatial segmentation as an effect of the deprivation condition under the same food and water delivery schedule. This analysis of behavioral spatial organization serves as an example of SDBA under different motivational operations.

Each plot depicts the area of the modified open field chamber ($92 cm \times 92 cm$), with the location of both dispensers marked with red triangles (food at the top; water at the bottom), where commodities were delivered simultaneously. The location of the five generators is marked with black points, and the Voronoi diagram with five regions is differentiated by color. Each generator point defines a Region of Behavioral Relevance (RBR).

The first criterion for comparing the diagrams is the location of the generators and their proximity to the dispensers, with closer proximity indicating a stronger influence of the dispenser on the generator’s region.

The second criterion is the extension of the RBR, which is related to the relevance of each region given the multiplicative weight and location of the generators, with larger areas indicating greater behavioral relevance.

Finally, the type of arcs in each region illustrates the spatial influence between regions. Straight-line arcs indicate equal relevance and influence between adjacent regions, suggesting similar weights and distances. Curved arcs show a subordinate influence, where one region's generator has a significantly higher weight or closer proximity, resulting in a curved boundary due to the dominance in spatial influence.

The comparison of Voronoi diagrams was conducted both within subjects and between subjects. In the within-subject analysis, the difference in behavioral spatial organization between deprivation conditions in both subjects is noticeable. At the same time, the similarities between subjects under the different conditions are highlighted.

\subsubsection{Subject 1}

For Subject 1, under Food Deprivation (FD) condition, the five generators where close to the food dispenser wall. Three of them were near to the dispenser, two were actually adjacent to it, and two were at the corners. Regarding the extension of RBRs, one of the generators adjacent to the dispenser was by far the most extended. The arcs of the regions were circular, showing a significant influence and dominance of a regions close to the dispenser over the rest of the regions. It is noticeable that two of the five generators were located in the corners close to the dispenser, given that these locations usually have ecological relevance as a home base or safety zone for the Wistar rats. 

Under Water Deprivation (WD) condition, two of the five generators were located close to the water dispenser. Nevertheless, the remaining three generators were close to the food dispenser wall, with two close to the food dispenser and one at the corner. Regarding the extension of the RBRs, the largest was associated with a generator close to the food dispenser, while the second and third largest were associated to the water dispenser. The arcs of the RBRs were curved, showing a dominance of the main RBRs over the rest. 

Under Food and Water Deprivation (FWD) condition, four generators were located close to the food dispenser wall, and only one adjacent to the water dispenser. Regarding to the RBRs extension, the two largest regions were adjacent to the food dispenser, the third in size to the water dispenser, and the last two were associated with a corner close to food dispenser. 

Under No Deprivation (ND) condition, again the five generators were close to the food dispenser, but this time the most extended region was associated with a corner. It is noticeable that among $100$ posible regions, there was consistency in the location of generators, and only eight of them were among the five main regions. On the other hand, under all deprivation conditions, the regions adjacent to the food dispenser were by far the most relevant.  Only under WD were the regions adjacent to the water dispenser relevant, but not as much as the regions adjacent to the food dispenser. Under FWD, a lower relevance of the water dispenser emerge. Finally, under ND, the most relavant region was a corner, followed by a region adjacent to the food dispenser. 

\subsubsection{Subject 5}

For Subject 5, under Food Deprivation Condition (FD), three of the five generators were located close to the food dispenser wall, two adjacent to it and one at the corner; additionally, two generators were located close to the opposite wall at the corner near the water dispenser. Regarding the extension of the RBRs, the largest by far was an adjacent RBR to the food dispenser. Noticeably, the RBRs near the water dispenser together form a significant region that could function as a home base.

Under Water Deprivation Condition (WD), three of the five generators were close to the wall of the water dispenser and two close to the wall of the food dispenser. Regarding the extension of the RBRs, it is highlighted that the RBR adjacent to the water dispenser is by far larger than the RBR adjacent to the food dispenser. Nevertheless, the largest RBR was associated with a corner near the food dispenser, and another large RBR was associated with a corner close to the water dispenser. The arcs of the RBRs associated with the generators at the corners tended to be straighter, indicating that both had a similar influence and weight, so they could function as a double home base. In this diagram, the RBRs were significantly more distributed and less concentrated than under the FD condition.

Under Food and Water Deprivation (FWD), the generators were located in a similar way to the FD condition: three generators close to the food dispenser wall and two in a corner close to the water dispenser. Nevertheless, the RBRs associated with the corner increased in comparison to the FD condition. One of the regions adjacent to the food dispenser was by far the largest and most relevant, with curved arcs indicating subordination of the rest of the regions to it.

Under No Deprivation (ND) condition, four of the generators were located close to the food dispenser wall, three of them adjacent to the dispenser and one at the corner. Finally, a generator was located at a corner near the water dispenser. In contrast with the FD and FWD conditions, by far the largest RBR was associated with a corner. The curved arcs show that all regions were subsumed by the largest region; nevertheless, the influence of this largest region is not as significant as those associated with the food dispensers under FD and FWD.

The similarity of the patterns between subjects, as shown in the Voronoi diagrams for each condition, is noticeable. First, under FD, the most salient RBR by far is adjacent to the food dispenser. Second, the salience of RBRs associated with the food dispenser was always present, even under the WD and ND conditions. Third, the RBRs adjacent to the water dispenser were salient only under the WD condition. Fourth, under all conditions, there was always at least one RBR associated with a corner, functioning as a home base, but only under ND were such RBRs more salient than the rest. On the other hand, it is highlighted that the different spatial segmentation is an effect of the type of deprivation, under the same reinforcement schedule.

\section{Conclusions}

Each region in the weighted Voronoi diagram defines an RBR. The various organizations of these regions represent well-differentiated patterns of behavioral segmentation within the same experimental space, even under the same reinforcement schedule. In this work, they depict distinctive spatial behavioral patterns related to the deprivation conditions. These different spatial organizations suggest that the different deprivation conditions functioned as different Motivational Operations,  giving varying importance to the regions close to the dispenser and corners. Identifying these RBRs provides valuable information about behavior dynamics in relation to the deprivation conditions.

The weighted Voronoi diagrams differed across deprivation conditions, allowing for the differentiation of spatial patterns known as RBRs. These patterns show how deprivation conditions affect the spatial behavior associated with seeking and contacting food and water, offering crucial insights into a subject's behavioral adjustments under different motivational operations. One of the most impactful insights is the asymmetrical effect of deprivation conditions on behavioral spatial organization; that is, the spatial organization under Food Deprivation is not simply inverted or symmetrical in relation to the spatial organization under Water Deprivation. In addition, when the rats were exposed to both Food and Water Deprivation, there was no symmetrical distribution of the RBRs between regions close to both dispensers; instead, the RBRs near the food dispenser were significantly more salient. Finally, under No Deprivation, only the regions close to the water dispenser were irrelevant. These asymmetrical effects of deprivation conditions on the spatial organization of behavior, particularly in regions where commodities were delivered, suggest that deprivation conditions function differently and asymmetrically as Motivational Operations \cite{Lewon_2019,michael_establishing_1993}.

Voronoi diagrams have demonstrated their value in enriching the understanding of spatial variables and features related to different behavioral phenomena. For example, they have been used to study the spatial organization of behavior under different spatiotemporal schedules \cite{Avendano2024}. This study extends the range of applications to include Motivational Operations under the same spatiotemporal reinforcement schedules. Therefore, the present study provides a more comprehensive perspective on the application of Voronoi diagrams to different behavioral phenomena. Nevertheless, there is work to do in order to expand the application of this tool, such as extending the proposed approach to the analysis of other relevant behavioral phenomena, such as the spatial organization under aversive stimulation. 

Finally, there are several aspects for future work regarding the mathematical developments to extend the application of Voronoi diagrams to SDBA. Among them, the first is to analyze and compare different methods to weight the generator points. A second aspect is to analyze the evolution of the spatial organization of behavior over time using the evolution of Voronoi diagrams.

\section*{Acknowledgements} 
BZMP and VQ would like to thank the Consejo Nacional de Humanidades, Ciencias y Tecnologías (CONAHCYT) for the grants CVU1314783 and CVU1000927 awarded to pursue a Master's Degree in Mathematics and a PhD in Behavioral Sciences, respectively, at the Universidad Veracruzana.

\bibliographystyle{plain}
\bibliography{biblio.bib}



{\vskip 12pt}
\noindent
\footnotesize {\textit{
Corresponding authors are\\
Porfirio Toledo, ptoledo@uv.mx, \\
and Alejandro León, aleleon@uv.mx.}}

\end{document}